\documentstyle[12pt]{article}
\newcommand{\fig}[1]{Fig.\ref{#1}}
\newcommand{\nn}{\nonumber}
\newcommand{\lb}{\linebreak}
\newcommand{\nlb}{\nolinebreak}
\newcommand{\ns}{\normalsize}
\newcommand{\bc}{\begin{center}}
\newcommand{\ec}{\end{center}}
\newcommand{\bq}{\begin{equation}}
\newcommand{\eq}{\end{equation}}
\newcommand{\bqa}{\begin{eqnarray}}
\newcommand{\eqa}{\end{eqnarray}}
\newcommand{\ben}{\begin{enumerate}}
\newcommand{\een}{\end{enumerate}}
\newcommand{\eqn}[1]{Eq.(\ref{#1})}

\def\a2n{${\cal A}(2\nolinebreak \to n)$}
\def\l{\lambda}
\def\ll{{\lambda\over 16 \pi^2}}\def\L{{\cal L}}\def\F{{\cal F}}
\def\e{{1\over \epsilon}}\def\ep{\epsilon}\def\po{\phi_0}
\def\dt{{d^2\over d\tau^2}}\def\k{\vec{k}}\def\oh{{1\over 2}}
\def\pod{{d\over d\tau}\phi_0}\def\m{{\mu\over m}}
\def\logm{\log({4\pi\mu^2\over m^2})}\def\z{\zeta}
\def\PT{perturbation theory}\def\om{\omega}
\def\dx{{d\over dx}}\def\dxb{{d^2\over dx^2}}
\def\ev{\varepsilon}

\begin{document}
\input feynman
\pagestyle{empty}

\begin{flushright}CERN-TH.6852/93\\
\end{flushright}
\vspace*{1cm}
\begin{center}\begin{Large}
{\bf Multiscalar amplitudes to all orders
in perturbation theory}\end{Large}

\vspace*{2cm}

E.N.~Argyres   \\
{\ns Institute of Nuclear Physics, NRCPS
`$\Delta  \eta  \mu  \acute{o}  \kappa  \varrho
  \iota  \tau  o  \varsigma$', Greece.}\\
\vspace{\baselineskip}
Ronald~H.P.~Kleiss,\\
{\ns NIKEHF-H, Amsterdam, the Netherlands}\\
\vspace{\baselineskip}
Costas~G.~Papadopoulos\\
{\ns TH Division, CERN, Geneva, Switzerland}\\
%end{Large}
\vspace*{3cm}
Abstract\\[24pt] \ec
A method for calculating loop amplitudes at the multiboson
threshold is presented, based on Feynman-diagram techniques.
We explicitly calculate the one-loop amplitudes in both
$\phi^4$-symmetric and broken symmetry cases, using dimensional
regularization. We argue that, to all orders in the perturbation
expansion, the unitarity-violating behaviour
of the tree-order amplitudes persists.
\vspace{1cm}
\begin{flushleft} CERN-TH.6852/93\\March 1993\\
\vfill
\end{flushleft}

\newpage
\pagestyle{plain}
\setcounter{page}{1}

\par The high-multiplicity limit of processes involving scalar
particles has recently come under investigation
\cite{volo:1,akpa:1}. At tree order,
the cross section grows as $n!$, where $n$ is the number of
produced scalar particles, and it violates the unitarity bound
for sufficiently high energies. The construction of an effective
potential that respects the unitarity bound is possible
\cite{akpa:2}. In order to make this effective theory
consistent at any $n$ and for any energy,
derivative coupling must be added, which generally leads to a complicated
theory.
\par One-loop amplitudes have been calculated \cite{volo:2,smit:1}
recently. The method used in these calculations is based on the
fact that the generating function of one-loop threshold
amplitudes is obtained
by expanding the field around the classical background (which
is the generating function of tree-order threshold amplitudes) and
keeping the first term \cite{brow}. More recently Voloshin \cite{volo:3}
argued that in broken symmetry
$\phi^4$ theory, the factorial growth of the amplitude,
which is related to the
radius of convergence of the generating function,
persists to all orders in perturbative expansion.
In this letter we develop a method for calculating loop
amplitudes at threshold based on Feynman diagrams, in analogy
with the tree-order calculations \cite{akpa:1}. We start with
the $\phi^4$ theory. The Lagrangian is given by
\bq \L={m^2\over 2}\phi^2+{\l\over 24}\phi^4\;\;.   \eq
The
recursion relation for the one-loop amplitude is represented
diagrammatically in \fig{fi01}. It is of the form

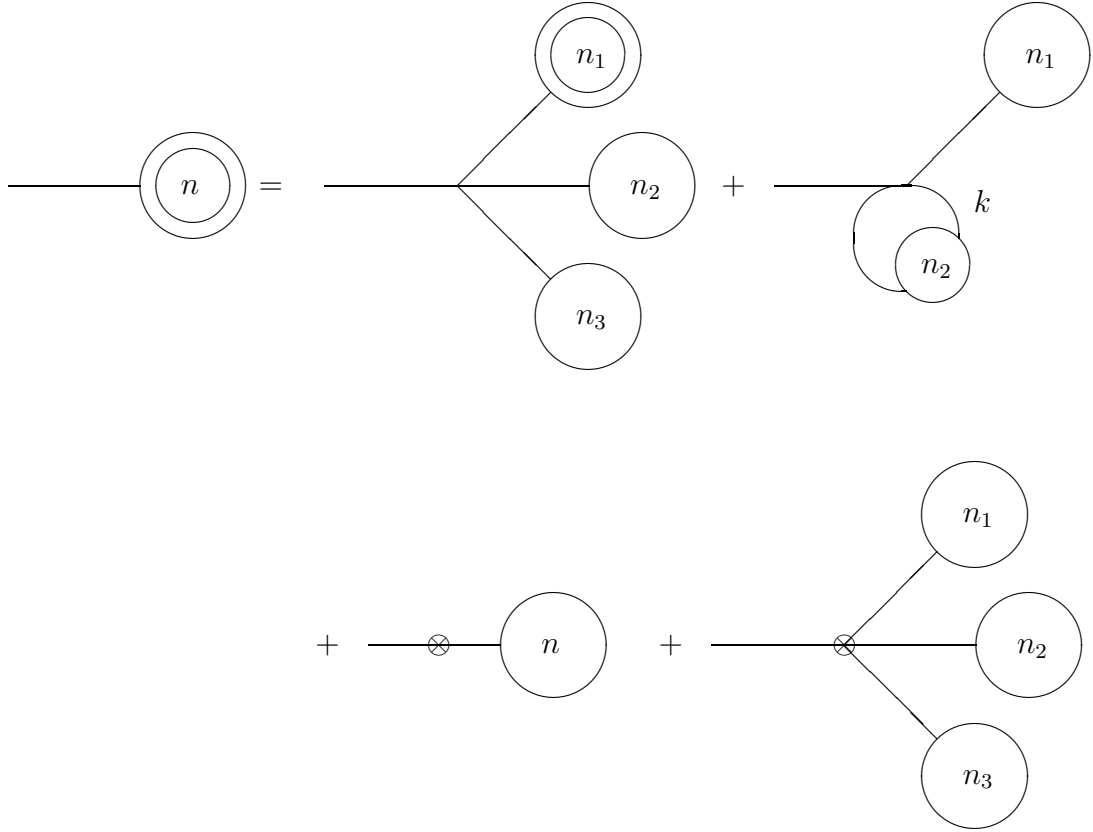
\begin{figure}[th]
\begin{picture}(20000,16000)
\drawline\fermion[\E\REG](0,8000)[5000]
\global\advance\pbackx by 2000
\put(\pbackx,\pbacky){\circle{4000}}
\put(\pbackx,\pbacky){\circle{3000}}
\global\advance\pbacky by -300
\global\advance\pbackx by -450
\put(\pbackx,\pbacky){$n$}
\global\advance\pbackx by 450
\global\advance\pbackx by 2500
\put(\pbackx,\pbacky){$=$}
\global\advance\pbackx by 2500
\global\advance\pbacky by 300
\drawline\fermion[\E\REG](\pbackx,\pbacky)[5000]

\drawline\fermion[\NE\REG](\pbackx,\pbacky)[5000]
\global\advance\pbackx by 1414
\global\advance\pbacky by 1414
\put(\pbackx,\pbacky){\circle{4000}}
\put(\pbackx,\pbacky){\circle{3000}}
\global\advance\pbacky by -300
\global\advance\pbackx by -450
\put(\pbackx,\pbacky){$n_1$}
\global\advance\pbackx by 450

\drawline\fermion[\E\REG](\pfrontx,\pfronty)[5000]
\global\advance\pbackx by 2000
\put(\pbackx,\pbacky){\circle{4000}}
\global\advance\pbacky by -300
\global\advance\pbackx by -450
\put(\pbackx,\pbacky){$n_2$}

\drawline\fermion[\SE\REG](\pfrontx,\pfronty)[5000]
\global\advance\pbackx by 1414
\global\advance\pbacky by -1414
\put(\pbackx,\pbacky){\circle{4000}}
\global\advance\pbacky by -300
\global\advance\pbackx by -450
\put(\pbackx,\pbacky){$n_3$}

\global\advance\pfrontx by 10000
\global\advance\pfronty by -300

\put(\pfrontx,\pfronty){$+$}

\global\advance\pfronty by 300
\global\advance\pfrontx by 2000

\drawline\fermion[\E\REG](\pfrontx,\pfronty)[5000]

\drawline\fermion[\NE\REG](\pbackx,\pbacky)[5000]
\global\advance\pbackx by 1414
\global\advance\pbacky by 1414
\put(\pbackx,\pbacky){\circle{4000}}
\global\advance\pbacky by -300
\global\advance\pbackx by -450
\put(\pbackx,\pbacky){$n_1$}

\global\advance\pfronty by -2000
\put(\pfrontx,\pfronty){\oval(4000,4000)[tr]}
\put(\pfrontx,\pfronty){\oval(4000,4000)[l]}
\global\advance\pfronty by -1000
\global\advance\pfrontx by 1000
\put(\pfrontx,\pfronty){\circle{2828}}
\global\advance\pfrontx by -450
\global\advance\pfronty by -300
\put(\pfrontx,\pfronty){$n_2$}

\global\advance\pfrontx by 2000
\global\advance\pfronty by 2300
\put(\pfrontx,\pfronty){$k$}
\end{picture}

\begin{center}
\begin{picture}(20000,16000)

\put(0,7700){$+$}
\drawline\fermion[\E\REG](2000,8000)[5000]
\global\advance\pmidx by -300
\global\advance\pmidy by -300
\put(\pmidx,\pmidy){$\otimes$}

\global\advance\pbackx by 2000

\put(\pbackx,\pbacky){\circle{4000}}
\global\advance\pbacky by -300
\global\advance\pbackx by -450
\put(\pbackx,\pbacky){$n$}
\global\advance\pbackx by 450

\global\advance\pbackx by 4000
\put(\pbackx,\pbacky){$+$}

\global\advance\pbackx by 2000
\global\advance\pbacky by 300

\drawline\fermion[\E\REG](\pbackx,\pbacky)[5000]

\global\advance\pbackx by -450
\global\advance\pbacky by -300
\put(\pbackx,\pbacky){$\otimes$}
\global\advance\pbackx by 450
\global\advance\pbacky by 300

\drawline\fermion[\NE\REG](\pbackx,\pbacky)[5000]
\global\advance\pbackx by 1414
\global\advance\pbacky by 1414
\put(\pbackx,\pbacky){\circle{4000}}
\global\advance\pbacky by -300
\global\advance\pbackx by -450
\put(\pbackx,\pbacky){$n_1$}

\drawline\fermion[\E\REG](\pfrontx,\pfronty)[5000]
\global\advance\pbackx by 2000
\put(\pbackx,\pbacky){\circle{4000}}
\global\advance\pbacky by -300
\global\advance\pbackx by -450
\put(\pbackx,\pbacky){$n_2$}

\drawline\fermion[\SE\REG](\pfrontx,\pfronty)[5000]
\global\advance\pbackx by 1414
\global\advance\pbacky by -1414
\put(\pbackx,\pbacky){\circle{4000}}
\global\advance\pbacky by -300
\global\advance\pbackx by -450
\put(\pbackx,\pbacky){$n_3$}

\end{picture}
\caption[.]{Diagrammatic representation of
the recursion formula for the one-loop amplitudes
${\cal A}(1\to n)$ (blob with two circles).
The blobs with on circle connected to a line
are the tree-order amplitude ${\cal A}(1\to n)$. The blob with one circle
connected to two lines
corresponds to the propagator with the emission of $n$ particles.}
\label{fi01}
\end{center}
\end{figure}

\bqa
{a_1(n)\over n!}&=&-i{\lambda\over 2}\sum
{i a_1(n_1)\over n_1!(n_1^2-1)}
{i a(n_2)\over n_2!(n_2^2-1)}
{i a(n_3)\over n_3!(n_3^2-1)} \nn\\
&&-i{\lambda\over 2}\mu^{2\ep}\sum
{i a(n_1)\over n_1!(n_1^2-1)}
\int {d^D k\over (2\pi)^D} {D(n_2;k)\over n_2!}\nn\\
&&-i T_2{i a(n)\over n!(n^2-1)}
-i {T_4\over 6}\sum
{i a(n_1)\over n_1!(n_1^2-1)}
{i a(n_2)\over n_2!(n_2^2-1)}
{i a(n_3)\over n_3!(n_3^2-1)}
\label{olre}\eqa
where $a_1(n)$ is the one-loop amplitude and $a(n)$ the
tree-order one. The integral over $k$ extends over the Minkowski space.
Throughout this paper we assume that $m=1$ and
we restore the mass dependence when it is necessary.
The counterterms contributions are
\bqa
T_2&=&\frac{1}{2}m^2\ll \biggl(\e+b\biggr)\\
T_4&=&\frac{3}{2}\lambda\ll \biggl(\e+c\biggr)
\eqa
where $b$ and $c$ are in principle,
arbitrary constants used to
define the renormalization conditions.
$D(n;k)$ is the propagator of the field with the
emission of $n$ particles \cite{volo:2}
(see \fig{fi02}). It satisfies the following recursion relation:
\begin{figure}[th]
\begin{center}
\begin{picture}(20000,16000)
\drawline\fermion[\E\REG](0,8000)[2000]

\global\advance\pbackx by 2000
\put(\pbackx,\pbacky){\circle{4000}}
\global\advance\pbackx by -450
\global\advance\pbacky by -300
\put(\pbackx,\pbacky){$n$}

\global\advance\pbackx by 2450
\global\advance\pbacky by  300

\drawline\fermion[\E\REG](\pbackx,\pbacky)[2000]

\global\advance\pbackx by 2500
\global\advance\pbacky by -300
\put(\pbackx,\pbacky){$=$}
\global\advance\pbacky by 300
\global\advance\pbackx by 2500

\drawline\fermion[\E\REG](\pbackx,\pbacky)[5000]

\drawline\fermion[\N\REG](\pbackx,\pbacky)[3000]
\global\advance\pbacky by 2000
\put(\pbackx,\pbacky){\circle{4000}}
\global\advance\pbackx by -450
\global\advance\pbacky by -300
\put(\pbackx,\pbacky){$n_1$}

\drawline\fermion[\S\REG](\pfrontx,\pfronty)[3000]
\global\advance\pbacky by -2000
\put(\pbackx,\pbacky){\circle{4000}}
\global\advance\pbackx by -450
\global\advance\pbacky by -300
\put(\pbackx,\pbacky){$n_2$}

\drawline\fermion[\E\REG](\pfrontx,\pfronty)[2000]

\global\advance\pbackx by 2000
\put(\pbackx,\pbacky){\circle{4000}}
\global\advance\pbackx by -450
\global\advance\pbacky by -300
\put(\pbackx,\pbacky){$n_3$}

\global\advance\pbackx by 2450
\global\advance\pbacky by  300

\drawline\fermion[\E\REG](\pbackx,\pbacky)[2000]

\end{picture}
\caption[.]{Diagrammatic representation of
the recursion relation for the propagator with
the emission of $n$ particles.}
\label{fi02}
\end{center}
\end{figure}
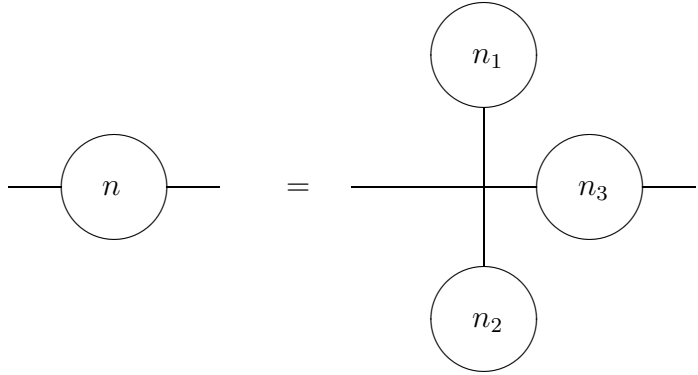

\bqa
{D(n;k)\over n!}={1\over[(k+nq)^2-1+i\ev]}
\;{\l\over 2}
\sum
{ia(n_1)\over n_1!(n_1^2-1)}
{ia(n_2)\over n_2!(n_2^2-1)}
{D(n_3;k)\over n_3!}\;\;.\eqa
Making the ansatz $D(n;k)=i n!d(n)$
and defining $g(x)=\sum d(n)x^n$,  we have
\bqa \Biggl(x^2\dxb+(2 k\cdot q +1)
x\dx+k^2-1+i\ev-{\l\over 2}\po^2\Biggr)g=1\eqa
where
\bq \po ={x\over 1-{\l\over 48}x^2} \eq
is the generating function of tree-order amplitudes \cite{beyo}.
Under the substitutions $x=\nlb i\sqrt{48/\l}e^\tau$, $u=e^\tau$,
$\om=\sqrt{\k^2+1-i\ev}$, $\ep=k\cdot q$ and
$g=\nlb e^{-\ep\tau}y(\tau)$ we obtain
\bqa \Biggl( \dt -\om^2 +{6\over\cosh^2(\tau)}\Biggr)y=e^{\ep\tau}
\;\;.\eqa
The solutions of the homogeneous part are
\bqa f_1&=&{(2-\om)(1-\om)+2 u^2(\om^2-4)+u^4(2+\om)(1+\om)\over
u^\om(1+u^2)^2}\nn\\
f_2 & = & f_1(\om\to -\om)\;\;.\eqa
The Wronskian is $W=2\om(\om^2-1)(\om^2-4)$ and
the solution for $g(\tau)$ is given by
\bq g(\tau)=-{e^{-\ep\tau}\over W}
\biggl( f_1(\tau)\int_{-\infty}^{\tau}
ds\;e^{\ep s}f_2(s)+f_2(\tau)\int_{\tau}^{\infty}ds\;e^{\ep s}
f_1(s)\biggr)\;\;.\eq
\par
Taking into account the results for $a(n)$ and $D(n;k)$ we
arrive at the following differential equation:
\bq
\biggl(\dt -1-{\lambda\over 2}\po^2\biggr)\phi_1 = -{\lambda\over 2}\po
\int {d^Dk\over (2\pi)^D} g(\tau;k)
+T_2\po+{T_4\over 6}\po^3
\label{dife}\eq
where $\phi_1$ is the generating function of 1-loop amplitudes,
$\phi_1=\sum b_1(n)x^n$ and  \lb $a_1(n)=\nlb -in!b_1(n)$.
The integral over $k$ now runs over the Euclidean space.
\par The integration over $k^0$ gives
\bqa
\mu^{2\ep}\int {d^Dk\over (2\pi)^D} g(\tau;k)=
-\mu^{2\ep}\int {d^dk\over (2\pi)^d} {f_1f_2\over W}
\eqa
where $d=D-1$.
It is straightforward to see that
\bq {f_1f_2\over W}={1\over 2(\k^2+1)^\oh}\biggl(1-
{\lambda\over 4}\po^2{1\over\k^2}+{\lambda^2\over 16}\po^4
{1\over \k^2(\k^2-3-i\ev)}\biggr)\;\;. \eq
\par In dimensional regularization, the first two terms
give poles $\e$, whereas the third term is finite, having
an imaginary part coming from the infrared singularity
at $\omega=2m$. The final result can be written as
\bq
\biggl(\dt -1-{\lambda\over 2}\po^2\biggr)\phi_1=B\po+C\po^3+F\po^5
\label{dif1}\eq
where
\bqa
B&=&{m^2\over 2}\ll\biggl(b+\gamma_E-1-\log({4\pi\mu^2\over m^2})
\biggr)\\
C&=&{\lambda\over 4}\ll\biggl(c+\gamma_E-2-\log({4\pi\mu^2\over m^2})
\biggr)\\
F&=&-{\lambda^3\sqrt{3}\over 1536\pi^2}
\biggl(\log{2+\sqrt{3}\over 2-\sqrt{3}} -i\pi\biggr)
\eqa
and $\gamma_E$ is the Euler-Mascheroni constant.
Assuming, for instance, that $B=0$, we find that the general
solution to \eqn{dif1} is given by
\bq \phi_1={3C\over\l}\biggl(\pod-\po\biggr)
+F{\l^3\over 24}{x^5\over(1-{\l\over 48}x^2)^3}
\;\;\;\;.\label{solu}\eq
The term multiplying $F$ has also been given
in ref.\cite{volo:2}.
\par We can repeat the same analysis in the case of the spontaneously
broken $\phi^4$ theory. The Lagrangian, after shifting,
is given by
\bq \L={m^2\over 2}\phi^2+{m\sqrt{3\l}\over 6}\phi^3+{\l\over 24}\phi^4
\eq
and the one-loop generating function satisfies the equation
\bqa
\biggl(\dt -1-\sqrt{3\l}\po{\lambda\over 2}\po^2\biggr)\phi_1
&=&-\biggl({\sqrt{3\l}\over 2}+{\lambda\over 2}\po\biggr)
\int {d^Dk\over (2\pi)^D} g(\tau;k)\nn\\
&&+T_1+T_2\po+{T_3\over 2}\po^2+{T_4\over 6}\po^3 \;\;.
\label{dif2}\eqa
The counterterms are given by
\bqa
T_1&=&{\sqrt{3\l}m^3\over 32 \pi^2}\biggl(\e+c_1\biggr)\nn\\
T_2&=&{\l m^2\over 8\pi^2}\biggl(\e+c_2\biggr) \nn\\
T_3&=&{3\l m\sqrt{3\l}\over 32\pi^2}\biggl(\e+c_3\biggr)\nn\\
T_4&=&{3\l^2\over 32\pi^2}\biggl(\e+c_4\biggr) \eqa
Performing the $k^0$ integration we get
\bq
\mu^{2\ep}\int {d^Dk\over (2\pi)^D} g(\tau;k)=
-\mu^{2\ep}\int {d^dk\over (2\pi)^d} {f_1f_2\over W}
\eq
where
\bq {f_1f_2\over W}={1\over 2(\k^2+1)^\oh}\biggl(1+
{3u\over(1+u)^2}{1\over (\k^2+{3\over 4})}+{9u^2\over(1+u)^4}
{1\over (\k^2+{3\over 4}) \k^2}\biggr) \;\;.\eq
Note that we can take the limit $i\ev\to 0$: this means that
the amplitude does not develop an imaginary part, as pointed out in
refs. \cite{beyo,smit:1}.
\par The final result is now written in the form
\bqa
\biggl(\dt -1-\sqrt{3\l}\po{\lambda\over 2}\po^2\biggr)\phi_1
&=&\biggl({\sqrt{3\l}\over 2}+{\lambda\over 2}\po\biggr)
{\sqrt{3}\over 2\pi}{u^2\over (1+u)^4}\nn\\&&+
C_1+C_2\po+C_3\po^2+C_4\po^3\eqa
where
\bqa
C_1&=&{\sqrt{3\l}\over 32\pi^2}\biggl(c_1+\gamma_E-\logm-1\biggr)\nn\\
C_2&=&{\l\over 16\pi^2}\biggl(2[c_2+\gamma_E-\logm]-{7\over 2}+
{\sqrt{3}\over2}\pi\biggr)\nn\\
C_3&=&{3\l\sqrt{3\l}\over 64\pi^2}\biggl(c_3+\gamma_E-\logm-2+
{\sqrt{3}\over3}\pi\biggr)\nn\\
C_4&=&{\l^2\over 64\pi^2}\biggl(c_4+\gamma_E-\logm-2+
{\sqrt{3}\over3}\pi\biggr)\eqa
Taking $c_i$ such that $C_i=0$, we find that
\bqa \phi_1={\sqrt{\l}\over 2\pi}{u^2\over(1+u^3)}\eqa
in agreement with \cite{smit:1}.
\par The previous calculations can be extended to higher orders
in \PT. In order to proceed, we have to add recursively
new counterterms to the original Lagrangian, in such a way that
we keep its form invariant. This can be done
(say for the $\phi^4$-symmetric case) by expressing, as usual,
the bare constants in terms of the renormalized ones:
\bqa
\l_0&=&\mu^{2\ep}\biggl( c_0(\l,\m,\ep)+\sum_{k=1}^{\infty}
{c_k(\l,\m)\over \ep^k}\biggr)\nn\\
m_0^2&=&m^2\biggl( b_0(\l,\m,\ep)+\sum_{k=1}^{\infty}
{b_k(\l,\m)\over \ep^k}\biggr)\nn\\
Z_\phi&=&\biggl( \z_0(\l,\m,\ep)+\sum_{k=1}^{\infty}
{\z_k(\l,\m)\over \ep^k}\biggr)\eqa
where $b_0$, $c_0$, $\z_0$ are analytic as $\ep\to 0$.
\par Let $a_L(n)$ being the $L^{\mathrm{th}}$-order amplitude for the
production of $n$ scalars at threshold and $\phi_L$ the
corresponding generating function. The equation it
satisfies is given by
\bq
\biggl(\dt -1-{\lambda\over 2}\po^2\biggr)\phi_L=
\l\phi_{L-1}\phi_1\po+\ldots+\F
+T^{(L)}_2\po+{T^{(L)}_4\over 6}\po^3
\label{difl}\eq
where $T^{(L)}_2$ and $T^{(L)}_4$ are the corresponding
$L^{\mathrm{th}}$-order
counterterms, whose finite part as $\ep\to 0$ is
in general arbitrary (renormalization-scheme-dependent)
and corresponds to the definition
of the $L^{\mathrm{th}}$-order
${\cal A}(1\to 1)$ and ${\cal A}(1\to 3)$ amplitudes.
The ellipsis
corresponds to lower-loop contributions and $\F$ represents
all explicit $L$-loop graphs (in analogy to the one-loop case).
\par Assuming now that the high-$n$ behaviour is
independent of the
renormalization prescription, i.e. the way we define the
parameters $b_0$, $c_0$ and $\z_0$ in the counterterm Lagrangian, it
is easy to see
that the leading behaviour of the solution $\phi_L$ should be
more singular at $\tau=\pm i{\pi\over 2}\;(x\equiv x_0=
\sqrt{{48\over \lambda}})$
than the combination $\pod-\po$ (see \eqn{solu}),
since we can compensate any change in the coefficient multiplying
$\po^3$ in \eqn{difl}, by adding
to the solution $\phi_L$ a term proporional to $\pod-\po$.
In general equating the most singular terms in \eqn{difl}
at $x=x_0$, we have that
\bq \rho(L)+2=\rho(L-1)+\rho(1)+1\;\;\;, L\ge 2\label{sing}\eq
where $-\rho(L)$ is the singularity-exponent at $x_0$. Barring
accidental cancellations, whose origin should be traced in a rather
non-trivial symmetry, a solution to \eqn{sing}
is of the form $\rho(L)=2L+1$.
Using the above result we obtain that, to
all orders in \PT\, the large-$n$ behaviour, which is
connected with the radius of convergence of the generating function,
is given by $\sim n!(n^2\l)^L$ for both the unbroken and
the broken symmetry cases. We find ourselves in agreement with
\cite{volo:3}, where the same conclusion has been drawn but from
a different point of view.
\par We conclude that the amplitudes ${\cal A}(H^*\to nH)$
exhibit a high-$n$ behaviour, which leads to unitarity violation
for all loops. Still, there is the possibility
that summing up all the perturbation series
one gets a consistent result.
Nevertheless, if this happens, it relies on
an essentially non-perturbative phenomenon. Having proven
that, for scalar amplitudes, \PT\ breaks down in the large-$n$ limit,
we can speculate that either a non-perturbative
phenomenon takes place at high energies, leading for instance
to a redefinition of the multi-Higgs states, in analogy with
the QCD confinement, or a symmetry should exist which
kills the factorial behaviour at threshold and hopefully
generates unitarity-respecting cross sections,
in analogy with gauge-symmetries and tree-order high-energy behaviour
\cite{tikt}.

\end{document}